\documentclass[12pt,preprint]{aastex}

\slugcomment{}

\shorttitle{Fluorine in AGB stars}
\shortauthors{Abia et al.}

\begin{document}


\title{Fluorine Abundances in Galactic AGB stars}


\author{C. Abia\altaffilmark{1}}
\affil{Dpto. F\'\i sica Te\'orica y del Cosmos, Universidad de Granada, 18071 Granada, Spain}
\email{cabia@ugr.es}

\author{K. Cunha\altaffilmark{2}}
\affil{National Optical Astronomy Observatory, P.O. Box 26732, Tucson, AZ 85726, USA}

\author{S. Cristallo\altaffilmark{1}}
\affil{Dpto. F\'\i sica Te\'orica y del Cosmos, Universidad de Granada, 18071 Granada, Spain}

\author{P. de Laverny\altaffilmark{3}}
\affil{University of Nice-Sophia Antipolis, CNRS (UMR 6202), Cassiop\'ee, 
Observatoire de la C\^ote d'Azur, B.P. 4229, 06304 Nice Cedex 4, France}

\author{I. Dom\'\i nguez\altaffilmark{1}}
\affil{Dpto. F\'\i sica Te\'orica y del Cosmos, Universidad de Granada, 18071 Granada, Spain}

\author{K. Eriksson\altaffilmark{4}}
\affil{Dept. of Physics \& Astronomy, Uppsala University, Box 515, 751 20 Uppsala, Sweden}

\author{L. Gialanella\altaffilmark{5}}
\affil{INFN Sezione di Napoli, Naples, Italy}

\author{K. Hinkle\altaffilmark{6}}
\affil{National Optical Astronomy Observatory, P.O. Box 26732, Tucson, AZ 85726, USA}

\author{G. Imbriani\altaffilmark{5}}
\affil{ Dipt. di Scienze Fisiche, Universit\'a Federico II, Naples, Italy}

\author{A. Recio-Blanco\altaffilmark{3}}
\affil{University of Nice-Sophia Antipolis, CNRS (UMR 6202), Cassiop\' ee, 
Observatoire de la C\^ote d'Azur, B.P. 4229, 06304 Nice Cedex 4, France}

\author{V.V. Smith\altaffilmark{6}}
\affil{National Optical Astronomy Observatory, P.O. Box 26732, Tucson, AZ 85726, USA}

\author{O. Straniero\altaffilmark{7}}
\affil{INAF-Osservatorio di Collurania, 64100 Teramo, Italy}

\and 

\author {R. Wahlin\altaffilmark{4}}
\affil{Dept. of Physics \& Astronomy, Uppsala University, Box 515, 751 20 Uppsala, Sweden}



\begin{abstract}
An analysis of the fluorine abundance in Galactic AGB carbon stars {\bf (24 N-type, 5 SC-type and 5 J-type)} 
is presented. This study
uses the state-of-the-art carbon rich atmosphere models and improved atomic and molecular
line lists in the $2.3~\mu$m region. F abundances significantly lower are obtained in comparison 
to previous study in the literature. The main reason of this difference is due to molecular
blends. In the case of carbon stars of SC-type, differences in the model atmospheres are also relevant. 
The new F enhancements are now in agreement with the most recent theoretical nucleosynthesis models 
in low-mass AGB stars, solving the long standing problem of F in Galactic AGB stars. Nevertheless,
some SC-type carbon stars still show larger F abundances than predicted by stellar models.
The possibility that these stars are of larger mass is briefly discussed.     

\end{abstract}



\section{Introduction}

The first  observational evidence of $^{19}$F stellar nucleosynthesis
was reported  by Jorissen, Smith  \& Lambert  (1992, hereafter
JSL). These authors derived F  enhancements up to a factor 50 solar
in a sample of Galactic AGB stars, and found
a  correlation  between  this  enhancement and the  C/O  ratio. Since the C/O  is 
expected to  increase as a consequence  of third
dredge  up  (TDU)  episodes  during  the  AGB  phase  (e.g.  Busso  et
al. 1999), this was interpreted as a clear evidence of F production in
these stars. Further observational  evidence of such a production exists
from studies of post-AGB stars (Werner, Rauch \& Kruk
2005)   and  planetary  nebulae   (Otsuka  et
al. 2008).  Other  sites for F  production have also been proposed:
Wolf-Rayet stars (Meynet \& Arnould
2000) and neutrino spallation  in core-collapse supernovae (Woosley \&
Haxton 1988). However, the role of these sources in the F
budget is  still uncertain  (Cunha et al.  2003; Palacios,  Arnould \&
Meynet 2005).  Nevertheless,  from Galactic chemical evolution models,
Renda et al. (2004) concluded that all three of these three sources are required
to explain the observed Galactic evolution of F, as deduced
from  abundance determinations in  field stars  (Cunha \&  Smith 2005;
Cunha, Smith \& Gibson 2008), although,  only in AGB stars
there exist an observational confirmation that F production is an
ongoing process.
 
Fluorine can be produced in AGB stars from the nuclear
chain $^{14}N(\alpha,\gamma)^{18}F(\beta^+)^{18}O(p,\alpha)^{15}N\\(\alpha,\gamma)^{19}F$;
where  protons are mainly  provided by $^{14}N(n,p)^{14}C$  and neutrons 
by  $^{13}C(\alpha,n)^{16}O$. However, the theoretical attempts made to explain the
JSL results are  unsatisfactory. The  problem  is that 
current AGB models fail  to
explain the highest F enhancements  found in the C-rich objects of the
JSL sample (Goriely \& Mowlavi 2000; Lugaro et
al.  2004).  This  discrepancy has  led  to  a  deep revision  of  the
uncertainties  in   the  nuclear   reaction  rates  involved   in  the
synthesis of F in AGB stars (Lugaro et al. 2004; Stancliffe
et  al.  2005),  to argue  for  alternative nuclear  chains and/or  to
propose  the  existence   of  non-standard  mixing/burning  processes,
similar to  those commonly used to  explain some of  the
isotopic anomalies found in dust grains formed in the envelopes of AGB
stars  (e.g. Nollett  et al.  2003; Busso  et al.  2007).  However, no
solution has been found up to  date, leaving the subject of the origin
of this element open.

Very  recently,  Abia et  al.  (2009,  hereafter  Paper I)  derived  F
abundances from  VLT spectra in  three Galactic AGB C-stars  in common
with  the JSL  sample,  namely: AQ  Sgr,  TX Psc  (N-type)  and R  Scl
(J-type), by using mainly the  HF R9 line at $\lambda 2.3358~\mu$m. In
that work, we showed that this HF  line is almost free of blends in AGB
C-stars  and thus is probably  the  best tool  for F  abundance
determinations  in  these  stars   (see also Uttenthaler et al. 2008). For these three stars, we derived  F
abundances  $\sim 0.7$  dex lower  in average  than those  obtained by
JSL. We ascribed this difference to molecular blends (mainly of C$_2$
and CN). These new F  abundances  are in   better  agreement  with the  most  recent
theoretical nucleosynthesis  predictions in low-mass  ($<3$ M$_\odot$)
TP-AGB models  (Cristallo et al.  2009). Motivated by this result, we have 
reanalysed  the  F abundances  in the  whole C-stars sample
studied by JSL using the same  tools as in Paper I.  In this Letter, we  confirm the results
found in Paper I and show  that the new F abundances nicely agree with the
theoretical  predictions of  low-mass TP-AGB  models.  This apparently
solves  the long  standing  problem  of F  enhancements  in AGB  stars
without requiring  the existence  of any extra  mixing/burning process.

\section{Analysis and Results}

Our data  consist of high-resolution spectra {\bf from $\sim 1.5$ to 2.5 $\mu$m}
obtained with the 4 m telescope at Kitt Peak Observatory and a Fourier
Transform  Spectrometer, which  have  been used  in several  abundance
analysis of AGB  stars (see Lambert et al.  1986 and JSL for details). Typical
resolutions ranged from 0.05 to 0.14 cm$^{-1}$ (R$=87000$ to 30000). These
spectra contain many HF lines  of the vibrational 1-0 band although we
used only  the lines R9  to R23. High resolution optical spectra (R$\sim  160000$) {\bf in the
range $\lambda 4600-8000$ {\AA}} obtained at
the  3.5  m  TNG  at  La  Palma  Observatory  with  SARG  echelle
spectrograph   were   also   analysed.   These   spectra   served  us to derive
the $s-$element content in {\bf the stars BL Ori, ST Cam, Y Tau, CY Cyg, RZ Peg and GP Ori
using the same procedure as in Abia et al. (2002)}. All program stars were ratioed
with hot stars to remove  telluric spectral features by using the task
{\it telluric} within  the IRAF package. The typical  S/N ratio of the
spectra  is  larger  than  100.   We adopted  the  stellar  parameters
(T$_{eff}$,    log    $g$,    [Fe/H],  $\xi$,   {\bf CNO abundances}, C/O  and
$^{12}$C/$^{13}$C  ratios)  derived   by  previous  studies:  Abia  et
al. (2002) and  Lambert et al. (1986) for N-type  stars, Abia \& Isern
(2000) for  J-type stars and Zamora  (2009) for stars  of SC-type (see
these works  for details).  In addition, we also  adopted the
$^{16}O/^{17}O/^{18}O$ ratios  from the literature when  available (Harris
et al.  1987), although this has  a minimal impact on  the derived F abundance. 
The method of analysis involves the comparison of theoretical
LTE spectra (computed with the {\it Turbospectrum} code) with  the  observed ones.   We  used  the  latest
generation of C-rich atmosphere  models for giant stars (Gustafsson et
al. 2008) and up-to-date molecular and atomic line lists in the spectral
ranges studied (see Paper I for details).

Figure  1 shows  an example  of comparison between  the observed  and
theoretical spectrum  for the C-star  UU Aur (N-type) in  two different
spectral  ranges.  In  particular, top
panel shows  the fit to the R15  and R16 HF lines, the sole lines used  by JSL to
derive the  F abundance in the  C-rich objects. It  can be seen that 
 we can  reproduce reasonably well the observed  spectrum and that
the synthetic spectrum  computed with the F abundance  obtained by JSL
(lower dashed line),  would   overestimate  the  F
abundance  in this star.  Our best  estimation in
this star  (log $\epsilon(F)=4.88$,  continuous line in  Fig. 1)  fits quite
well the  six HF lines used,  with a dispersion of  only $\pm0.04$ dex
(see also  Table 1)\footnote{{\bf We notice, that the F abundances derived in the stars
AQ Sgr and R Scl by using both CRIRES and FTS spectra are in perfect agreement.}}. 
Similar quality fits  were obtained for all the stars  except for 
V Hya, which was excluded from the analysis since its effective
temperature is outside the range covered  by our grid of models.  
Table 1 lists the final F abundance derived in each star, assigning double weight to the 
R9 line when estimating the mean F abundance because this is the less blended
among the available HF lines (see  Paper I for a discussion). For most
of  the stars,  more  than three  different  HF lines  were used.  The
agreement between  the different  lines is  in general good  with a
typical mean dispersion  of $\pm 0.08$ dex. The formal  error in the F
abundance due  to uncertainties in  the stellar parameters  ranges from
$0.20-0.30$ dex,  depending on the particular HF  line. Considering the
abundance dispersion and the uncertainty in the
continuum  position as additional sources of error, we  estimate an total uncertainty  of $\pm
0.30-0.35$  dex in  the [F/H]\footnote{The  solar F  abundance adopted
here is 4.56  (Grevesse, Asplund \& Sauval 2007).}  ratio (see Paper
I).   However, the  abundance ratio  between F  and any  other element
typically has a smaller error  since  some  of  the  above
uncertainties  cancel  out  when  deriving the  abundance  ratio: for the [F/Fe] 
ratio we estimate a total uncertainty of $\pm 0.25$ dex.

The  third  column in  Table  1 shows  the  difference  found between the  F
abundance  derived  by  JSL and  this  work.   The
differences range  from 0.3  dex up  to more than  1 dex,  the average value
being  $\Delta$log $\epsilon$(F)$=+0.7$, confirming  the main
result of Paper  I. As noted there, differences in  the stellar parameters adopted 
in the analysis cannot explain this discrepancy. 
Actually, the major differences are found for the T$_{eff}$
values, which for the N- and J-type stars in the sample, do not exceed
150  K. In the  case of  SC-type stars,  differences in  adopted T$_{eff}$
may  reach up to  300 K, but  in any case, this cannot explain the differences of  more than $\sim 1$ dex in the
F  abundance  found  in  some  of  these  stars.  Differences  in  the
atomic data are  also discarded  since we  used
exactly the same  HF line list than JSL.  Finally, differences between
the grid of model atmospheres might  account up to a maximum of 0.15 dex,
for the  N- and  J-type stars.  For the SC-type  stars, we found the largest abundance differences 
(see Table 1). For these stars, atmosphere models indeed play a significant  role.  JSL used for the analysis
of SC stars Johnson (1982) models, while we use those
from Gustafsson  et al. (2008). {\bf For instance, considering the typical stellar parameters of
a SC star in the sample (T$_{eff}$/log g/[Fe/H]$=3500/0.0/0.0$ and C/O$=1.0$), we find
differences in  the F abundance from $+0.4$ dex  (R9, R15 and R16 lines) to $+0.02$ dex (R23 line)} when 
using Johnson models instead of Gustafsson et al. ones, {\bf but it may reach up to $+0.6$ dex for lower T$_{eff}$}.  
In any case, for  the N-  and J-type  stars, 
we  conclude (as in Paper I) that the  difference between the 
molecular line  list used is the  main cause of  the discrepancy with respect to
the  analysis of  JSL {\bf(see Table 2)}. For  the SC-type
stars, it is a combination of that, the use of a different grid
of model atmospheres and, in the case of CY Cyg and GP Ori,
the  adoption here of  a  T$_{eff}\sim 300$  K  cooler.  It should  be
mentioned  that  the structure  of the  C-rich  atmosphere models
change dramatically with a tiny variation  of the C/O ratio when it is
very close  to 1, as it  is the case for  SC stars. Therefore,
the abundance  analysis   in  these  stars  are   affected  by  systematic
differences between the model atmosphere  and the real star and/or the
treatment  of the  molecular  equilibrium in  the  computation of  the
atmosphere models.  In consequence, for SC-type  stars the uncertainty
in [F/H] is certainly larger than $\pm 0.3$ dex.

\section{Discussion}
The main consequence of the new  F abundances is that
the large [F/Fe] (or [F/O]) ratios (up to 1.8 dex) found by JSL in
Galactic  AGB C-stars are  systematically reduced.  The  largest F
enhancements  are  now close to  $\sim  1$  dex.  These  enhancements  can  be
accounted  for by  current low-mass  TP-AGB nucleosynthesis  models of
solar metallicity, as  we will show below. As noted  in \S 1, during
the  ascension along the  AGB,  fresh carbon  is mixed within the
envelope due to TDU episodes. Eventually,  an O-rich AGB star
becomes  a  C-star  when  the   C/O  ratio  in  the  envelope  exceeds
1. Similarly, F is also expected  to increase in the envelope during 
the  AGB phase, thus  a fluorine vs. carbon correlation should  exist. Figure  2 shows  the observed
relationship derived  in this study.  We have also included in  this figure
the  intrinsic\footnote{Intrinsic  AGB   stars  have  their  envelopes
  polluted by nucleosynthesis products made {\it in situ}. Extrinsic  stars, 
on the contrary, own  their  chemical peculiarities to a  mass  transfer  episode  in a  binary
  system.  All  the  O-rich  stars  shown in  Fig.  2  show  $^{99}$Tc
  ($\tau_{1/2}\sim  2\times 10^5$  yr)  in their  envelopes (Smith  \&
  Lambert  1990), revealing their intrinsic nature.}  O-rich  AGB stars
studied  by  JSL (not  analysed  here, open circles). Excluding  the
J-type\footnote{The origin  of J-type  C-stars is still  unknown. They
  are  slightly less luminous  than N-type  C-stars and  show chemical
  peculiarities  rather  different  with respect to these stars  (e.g.  Abia  \&  Isern
  2000). It has  been suggested that they might  be the descendants of
  the early R-type stars, however  this has been questioned recently 
  (Zamora 2009).}  stars (triangles), a clear increase of the  F abundance with 
the C  abundance can be seen.  This
behaviour is  well reproduced by theoretical AGB models. Lines in Fig. 2 show 
the predicted F and C content in the envelope (Cristallo et al. 2010) for
a  1.5 M$_\odot$ with different metallicities (continuous  lines) and  2  M$_\odot$ model (dashed  line)
with solar metallicity. These metallicities match those of the stellar sample
($-0.5\leq$[Fe/H]$\leq  0.1$). The theoretical curves start  with an envelope composition
 as determined by the first dredge-up and end at the last TDU episode. 
On the other hand, Fig. 2 seems to indicate that O-rich stars and SC stars   
(squares) have F abundances systematically larger than N-type stars 
(dots) for a given C abundance. Similarly to the intrinsic C-stars, also
in the case  of the  O-rich stars this  
may be simply due to a systematic error  affecting the  analysis:  in fact, for stars
of spectral types  K and  M (with [Fe/H]$\sim  0.0$), JSL derived an 
average F abundance $\sim0.13$ dex higher (4.69) than the solar F abundance adopted
here (4.56). These K and M stars are sub-giants or RGB stars, thus they are not expected to present F
enhancements. Excluding that the Sun might be  
anomalous with respect to nearby stars of similar metallicity (F determinations in
unevolved dwarf stars are needed in this sense), blends may also play a role in the analysis of F lines
in O-rich AGB stars. Decreasing the  F abundances 
by 0.13 dex  in them, the agreement 
with theoretical predictions would be definitely better. However, it does not apply to the case of the
star BD$+48^o1187$ (at log $\epsilon$(F)$\sim  5.5$ in Fig. 2); the
large F enhancement in this star can be only obtained in metal-poor AGB models, but
this  star has [Fe/H]$=0.07$.  Something similar  seems to  occur with
SC-type stars. SC  stars are AGB stars with C/O$\approx 1$. According to
the accepted chemical (spectral) evolution along the AGB (M$\rightarrow$MS$\rightarrow$S$\rightarrow$SC$\rightarrow$N), 
these stars should  present  equal-or-slightly lower F enhancements 
than  N-type stars. However, some of the studied SC stars clearly deviate from this sequence, showing
larger F abundances with respect to N stars. This is reinforced by WZ Cas\footnote{This star  has also been
classified  as J-type.} ([Fe/H]$\sim  0.0$, at the  right upper
corner  of  Fig.  2), which shows a  huge  F
enhancement. We remind, however, that this  star is indeed peculiar: it  is one of the
few  super   Li-rich  C-stars  known,  also presenting   very  low
$^{12}$C/$^{13}$C  (4.5) and  $^{16}$O/$^{17}$O (400)  ratios.  The fact that most  of the SC-stars studied here 
show low or no $s-$element enhancements  (see below), makes the evolutionary status of these stars very uncertain.

The connection between the F and the $s$-process is less straightforward. There are two different 
contributions to the F production in low mass AGB stars. The first comes from the 
$^{15}$N production in the radiative $^{13}$C pocket, the site where the s-process main 
component is built up: when, during the interpulse period,  the $^{13}C(\alpha,n)^{16}O$ is activated within 
the pocket, some protons are released by the main poisoning reaction,  $^{14}N(n,p)^{14}C$, thus producing $^{15}$N via 
 $^{18}O(p,\alpha)^{15}N$. The second contribution involves the same reaction chain, but it is activated 
when the $^{13}$C left by the advancing H-burning shell is engulfed into the convective zone generated 
by a thermal pulse. In the latter case, the correlation with the $s-$process is less stringent, because 
the resulting neutron flux is not enough to give rise to a sizable production of elements beyond iron. 
At nearly solar metallicity, the latter contribution accounts for $\sim$70\% of the total F production {\bf(e.g. Cristallo 2006)}.  
Nevertheless, a correlation between the F and the $s-$element overabundance is in any case expected 
if a large enough $^{13}$C pocket forms after each dredge up episode. Figure  3 shows  the  new [F/Fe]  
ratios  in  C-stars vs.  the observed average $s-$element enhancement {\bf(from Abia et al. 2002; Zamora 2009 or
derived in this work, see \S 1 )}.  J-type stars are not shown in this figure as they do not  have $s-$element enhancements
(see  Table 1;  Abia \&  Isern  2000). Similarly  to Fig.  2, we  have
included  the  intrinsic O-rich  stars  analysed  by  JSL taken  their
$s-$element content from Smith \& Lambert (1990). It is clear that the
F and  $s-$nuclei overabundances  correlate and  that this
correlation  is  nicely reproduced by theoretical
TP-AGB stellar  models  (Cristallo et  al. 2010).  The
theoretical predictions in Fig. 3 correspond to 1.5, 2 and 3 M$_\odot$
TP-AGB  models   with  Z$=0.008$  ([Fe/H]$\sim   -0.25$,  the  average
metallicity of the stars analysed  here) starting from the 1$^{st}$ TP
pulse. The number  of TPs calculated until the  occurrence of the last
TDU is indicated in the figure for each  model. Models with a metallicity
within $\sim \pm 0.2$ dex  of this value do not significantly differ
from those in  Figure 3. In addition, with the combination of models with different stellar mass and
metallicity, 
it is possible not only to reproduce
the new observed [F/Fe]  vs. $s-$element relationship, but also
the  detailed  F  and  $s-$element  abundance pattern  observed  in  a
particular star (see e.g. Fig. 3 in  Paper I). Note that this is not possible for most
of the stars when adopting the F determinations of JSL.

As already noted, SC-type stars (except GP  Ori) show larger F abundances than
predicted by low-mass AGB models with the same $s-$element  enhancement. 
A possible explanation could be that the SC stars are more massive (in average) than the 
bulk of the N-type C-stars. From AGB models with mass of about 3-5 M$_\odot$, we expect smaller 
$^{13}$C pockets and weaker $s-$element surface enrichments, even if the number of TPs
and TDU episodes is larger. It occurs because the larger the core mass is, the smaller the He-rich intershell is 
with a steeper pressure gradient. Notwithstanding, the $^{13}$C left in the ashes of the H-shell still 
provide a substantial contribution to the fluorine production. Recent studies of the luminosity 
function of Galactic C-stars (Guandalini 2008) support such a hypothesis. They indicate that the SC-type 
stars are among the most luminous AGB C-stars. Once again, the case of WZ Cas is particularly extreme. 
This  star shows a huge F abundance ([F/Fe]$=+1.15$) with almost no $s-$element enhancement. Its large Li abundance
might be interpreted as a consequence of the hot bottom burning (HBB) mechanism which operates only
for M$\geq 5$ M$_\odot$. {\bf However, the operation of the HBB would be at odds with the
presence of F in the envelope since $^{19}F(p,\alpha)^{16}$O might destroy $^{19}$F}. Detailed abundance 
studies (including  F) in a larger sample of
SC-stars is needed to confirm this hypothesis.

\section{Summary}
Fluorine abundances in Galactic AGB carbon stars are determined from high resolution
infrared spectra using state-of-the-art C-rich spherical model atmospheres, updated 
atomic and molecular line lists and LTE spectral synthesis. We derive F abundances 
that are in average $\sim 0.7$ dex lower than those derived in previous studies. A clear
correlation between the F and $s-$element enhancements is found. The new
fluorine enhancements can be explained by standard low-mass AGB stellar models. F abundance determinations
 appear  as a valuable tool to test theoretical AGB models. 
Studies of the F abundance in metal poor unevolved
and AGB stars are mandatory to understand the Galactic abundance evolution of this element
and to evaluate the role of AGB stars in its origin, respectively.

\acknowledgments

Part   of   this   work   was   supported  by   the   Spanish   grants
AYA2008-04211-C02-02 and  FPA2008-03908 from  the MEC.  P.  de Laverny
and  A. Recio-Blanco  acknowledge the  financial support  of Programme
National  de Physique Stellaire  (PNPS) of  CNRS/INSU, France. KE gratefully
acknowledges support from the Swedish Research Council. We are
thankful  to B.  Plez for  providing us  molecular line  lists  in the
observed infrared domain.

{\it Facilities:} \facility{ORM TNG (SARG) and NOAO Kitt peak (FTS)}.

\clearpage

\begin{figure}
\epsscale{1.0}
\plotone{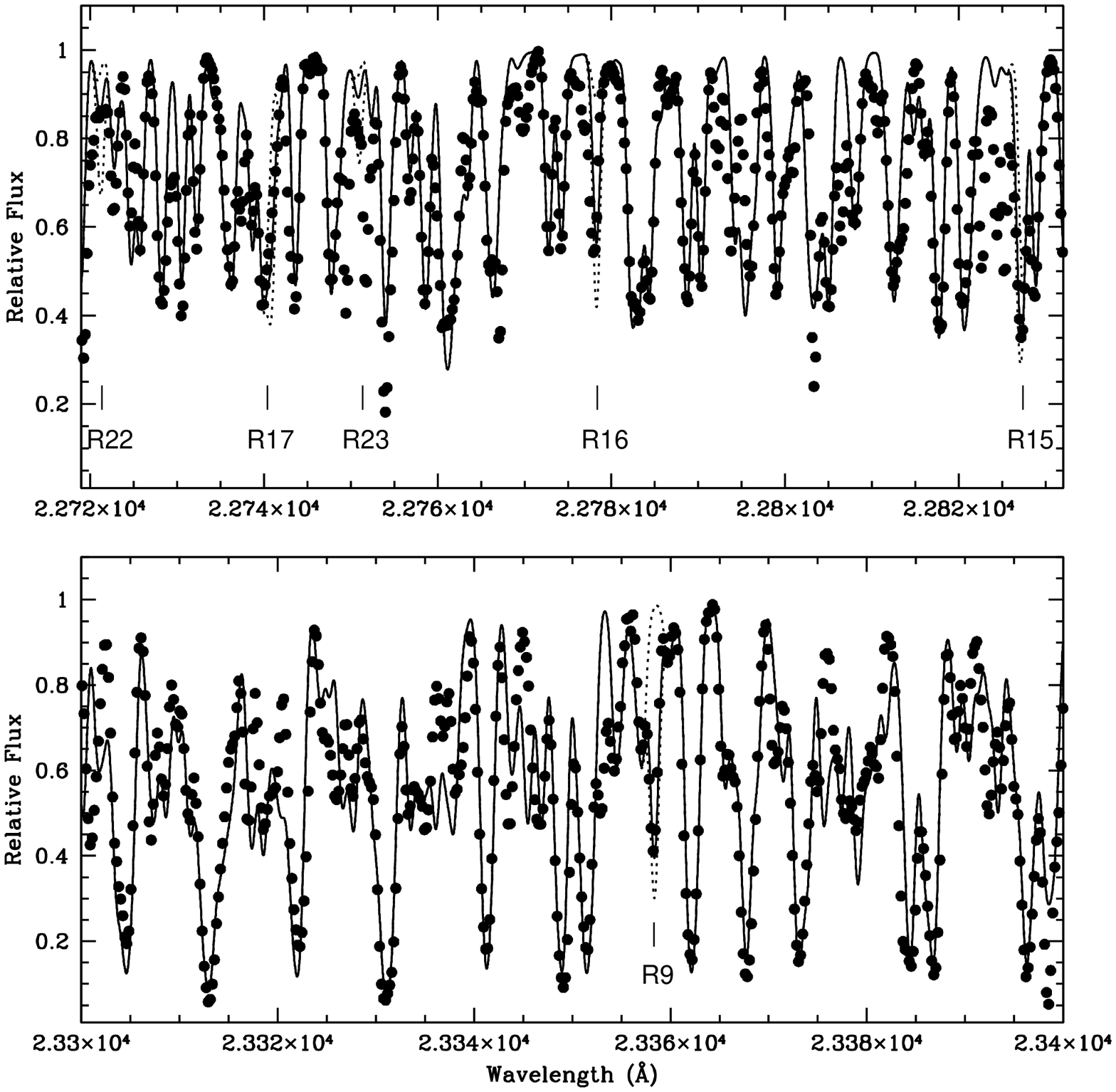}
\caption{Comparison of the observed and synthetic spectrum of UU Aur (dots)  with 
an identification of the available HF lines. Dashed lines represent the
synthetic spectra calculated with no F and with the abundance value obtained by JSL, respectively. 
The continuous line is the synthetic spectrum calculated
with the abundance given in Table 1, log $\epsilon$(F)$=4.88$.\label{fig1}}
\end{figure}

\clearpage
\begin{figure}
\epsscale{1.0}
\plotone{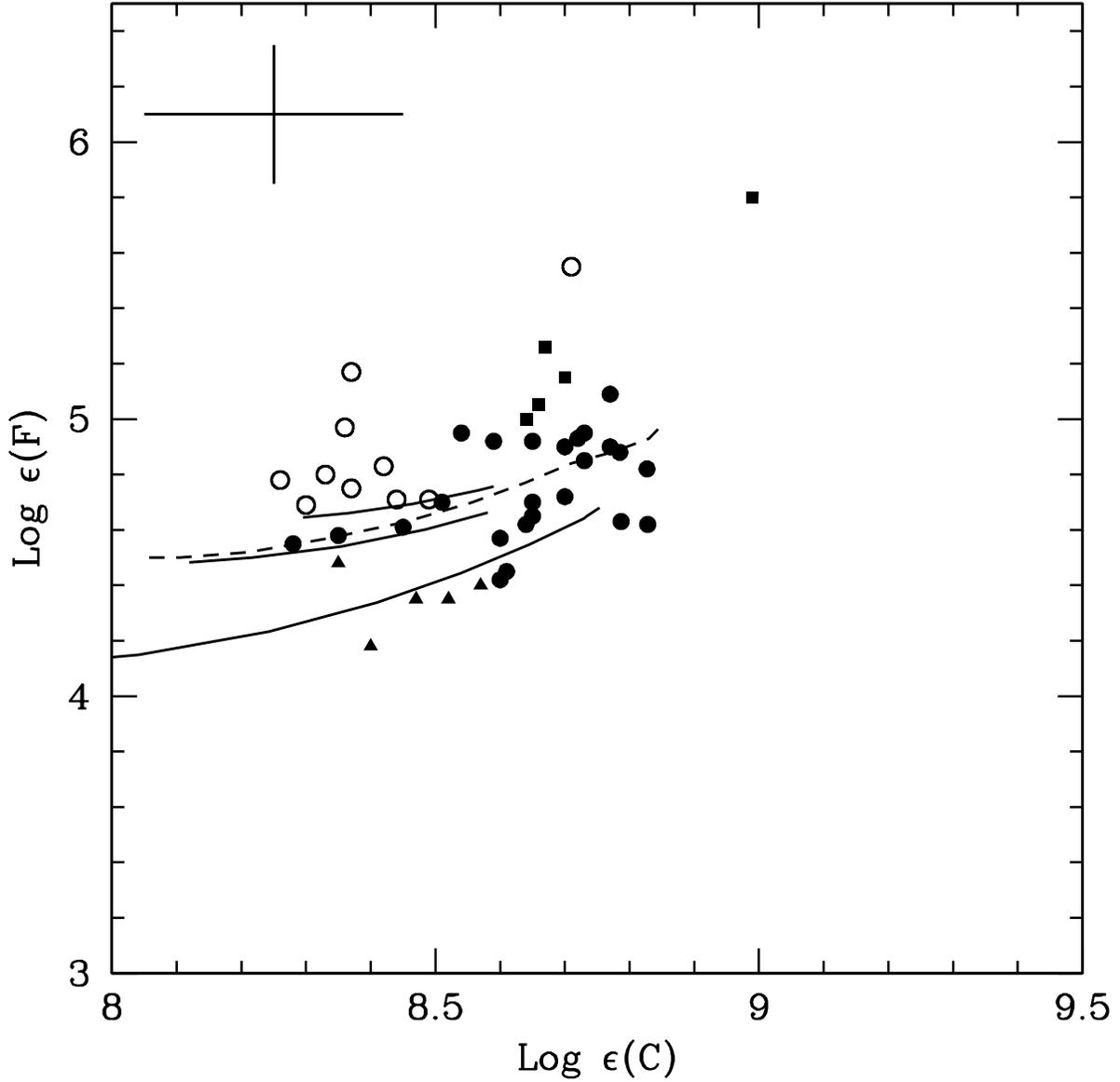}
\caption{ Logarithmic abundances of fluorine vs. carbon. Symbols: filled circles, N-stars; triangles, J-type; squares,
 SC-type; open circles, intrinsic O-rich AGB stars from JSL. Lines are theoretical predictions for a
1.5 M$_\odot$, TP-AGB model with metallicities $Z=0.02,~Z_\odot$ and 0.006 (continuous lines from up 
to down), and a 2 M$_\odot$, $Z=Z_\odot$ model (dashed line), respectively. \label{fig2}}
\end{figure}
\clearpage

\begin{figure}
\epsscale{1.0}
\plotone{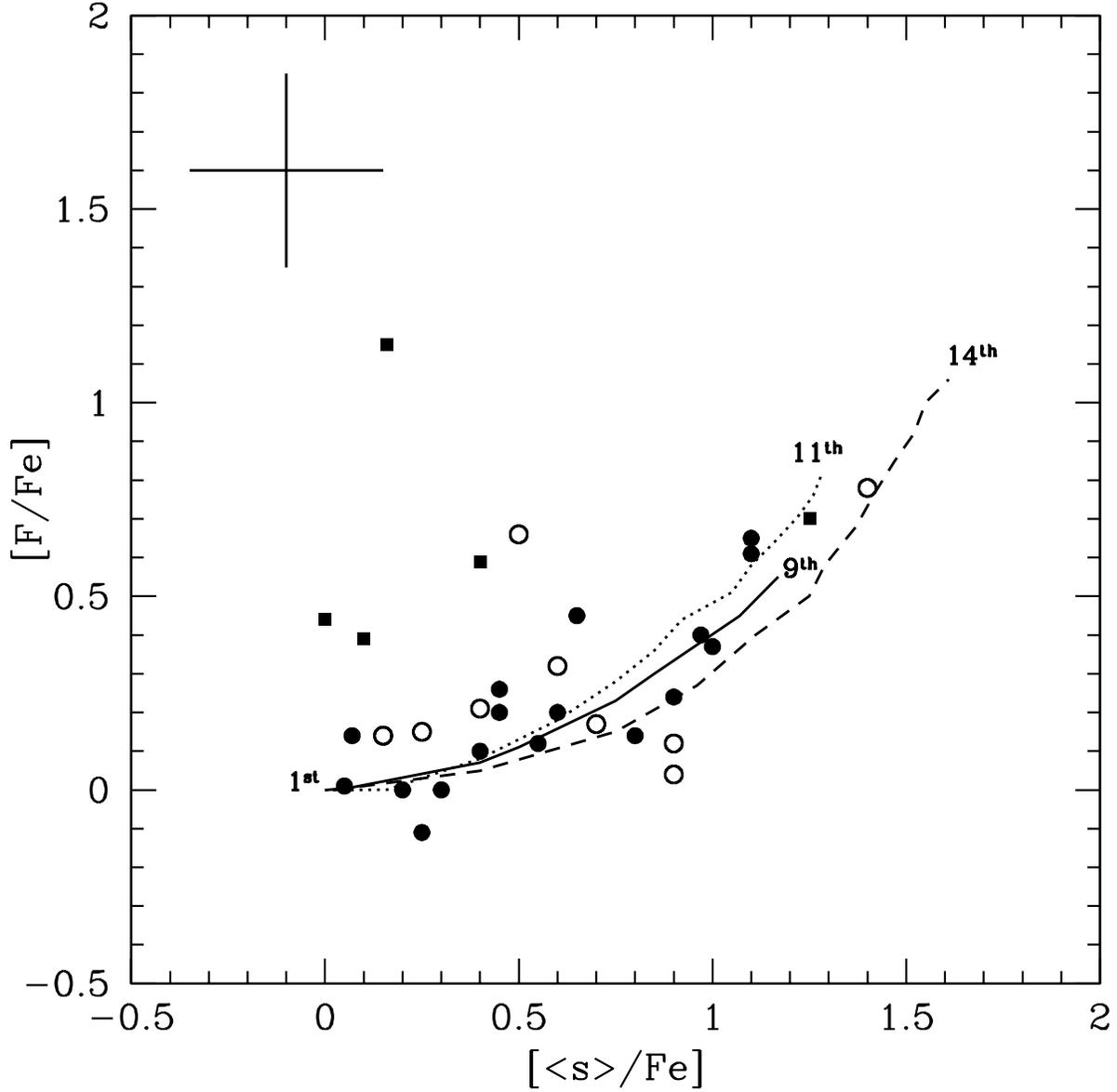}
\caption{Fluorine vs.  average s-element enhancements  in galactic AGB
  stars. Symbols as in Fig 2. Lines are theoretical predictions for 1.5, 2 and 3 M$_\odot$,
  $Z=0.008$   TP-AGB   models  (solid, dashed   and  dotted   lines,
  respectively) from Cristallo et al. (2010). The number of TPs achieved by each models is also indicated.\label{fig3}}
\end{figure}

\begin{deluxetable}{lccccc}
\tablecaption{Abundances.\label{tbl-1}}
\tablewidth{0pt}
\tablehead{
\colhead{Star} & \colhead{log $\epsilon$(F)\tablenotemark{a}}& \colhead{$\Delta$log $\epsilon$(F)\tablenotemark{b}} & \colhead{C/O} & \colhead{[F/Fe]}&
\colhead{[$<$s$>$/Fe]}\\
}
\startdata
$N~stars$ &  &   & &  &\\
AQ Sgr& 4.63$\pm0.02$ (4)& 0.85& 1.03& 0.10& \nodata\\
BL Ori& 4.92$\pm0.17$ (4)& 0.57& 1.04& 0.37& 1.00\\
R Lep & 4.45$\pm0.20$ (2)& 0.47& 1.03& 0.34& \nodata\\
RT Cap& 4.95$\pm0.01$ (5)& 0.36& 1.10& 0.40& \nodata\\
RV Cyg& 4.70$\pm0.10$ (4)& 0.63& 1.20 & 0.15& 0.07\\
S Sct & 4.62$\pm0.03$ (4)& 1.08& 1.07& 0.10& 0.40\\
SS Vir& 4.55$\pm0.07$ (2)& \nodata& 1.05& 0.00& 0.30\\
ST Cam& 4.72$\pm0.14$ (4)& 0.94& 1.14& 0.20& 0.45\\
TU Gem& 4.92$\pm0.07$ (5)& 0.88& 1.12& 0.36& \nodata\\
TW Oph& 4.61$\pm0.06$ (3)& 0.30& 1.20& 0.46&\nodata\\
TX Psc& 4.82$\pm0.10$ (6)& 0.73& 1.03& 0.65& 1.10\\
U Cam & 4.58$\pm0.07$ (4)& 0.54& 1.30& 0.12& 0.55\\
U Hya & 5.09$\pm0.08$ (5)&  \nodata& 1.05& 0.61& 1.10\\
UU Aur& 4.88$\pm0.04$ (5)& 0.76& 1.06& 0.26& 0.45\\
UX Dra& 4.85$\pm0.15$ (4)& 0.64& 1.05& 0.45& 0.65\\
V460 Cyg&4.65$\pm0.10$ (4)& 0.68& 1.06& 0.14& 0.80\\
V Aql & 4.62$\pm0.14$ (5)& 0.54& 1.25& 0.00& 0.20\\
VY UMa& 4.70$\pm0.08$ (3)& 1.02& 1.06& 0.24& 0.90\\
W CMa & 4.95$\pm0.08$ (6)& 0.56& 1.05& 0.20& 0.60\\  
W Ori & 4.42$\pm0.05$ (4)& 0.65& 1.16& -0.11& 0.25\\
X Cnc & 4.93$\pm0.10$ (3)& 0.76& 1.14& 0.67& \nodata\\
Y Hya & 4.90$\pm0.02$ (5)& 0.33& 1.52& 0.44& \nodata\\
Y Tau & 4.57$\pm0.09$ (6)& 0.49& 1.04& 0.01& 0.05\\
Z Psc & 4.90$\pm0.13$ (4)& 0.42& 1.01& 0.40& 0.97\\
$SC~stars$& & & & & \\
CY Cyg& 5.05$\pm0.05$ (4)& 0.98& 1.04& 0.39& 0.10\\
FU Mon& 5.15$\pm0.10$ (5)& 1.35&1.00& 0.59& 0.40\\
GP Ori& 5.26$\pm0.08$ (6)& 0.94&1.00& 0.70& 1.25\\
RZ Peg& 5.00$\pm0.10$ (6)& 1.06&1.00& 0.44& 0.00\\
WZ Cas& 5.80$\pm0.17$ (6)& 0.40&1.01& 1.15& 0.16\\
$J~stars$& & & & &\\
R Scl & 4.18$\pm0.03$ (3)& 1.22&1.34 & 0.09& \nodata\\
RY Dra& 4.48$\pm0.06$ (3)& 0.44& 1.18& -0.08& $<$0.20\\
T Lyr& 4.35$\pm0.00$ (2)& 0.45&1.29& -0.21& \nodata\\
VX And& 4.35 (1)& \nodata&1.76& -0.20& 0.2\\
Y Cvn&  4.40 (1)& 0.50&1.09& -0.10& $<$0.2\\
\enddata
\tablecomments{The abundances of fluorine are given using the definition
log $\epsilon$(X)$= 12 +$ log (X/H), where
(X/H) is the abundance of the element X by number.}
\tablenotetext{a}{The number in parenthesis indicates the number of HF lines used.}
\tablenotetext{b}{Abundance difference with respect to JSL.}
\end{deluxetable}

\begin{deluxetable}{lcc}
\tablecaption{Main blends affecting the HF lines.\label{tbl-2}}
\tablewidth{0pt}
\tablehead{
\colhead{HF line} & \colhead{Blend}& \colhead{$\Delta$[C/H]$=\pm 0.2$\tablenotemark{a}}\\
}
\startdata
R9  & \nodata                              & $\mp 0.01$ \\
R15 & $^{12}$C$^{14}$N, $\lambda~ 22827.354$  & $\mp 0.25$\\
R16 & $^{12}$C$^{12}$C, $\lambda~22778.775$  & $\mp 0.15$\\
R17 & $^{12}$C$^{14}$N, $\lambda~22740.700$  & $\mp 0.03$\\
R22 & $^{13}$C$^{14}$N, $\lambda~ 22721.294$  & $\mp 0.05$\\
R23 & $^{12}$C$^{14}$N, $\lambda~22751.140$ & $\mp 0.02$\\
\enddata
\tablenotetext{a}{Variation of the F abundance derived from each HF line due to changes in the carbon abundance 
for a representative case T$_{eff}$/log g/[Fe/H]$=2800/0.0/0.0$
and C/O$=1.06$.}   
\end{deluxetable}


\begin{thebibliography}{}
\bibitem[Abia (2000)]{abi00} Abia, C., \& Isern, J. 2000, ApJ, 536, 438
\bibitem[Abia et al. (2002)]{abi02} Abia, C., et al. 2002, ApJ, 578, 817
\bibitem[Abia et al. (2009)]{abi09} Abia, C., et al. 2009, ApJ, 694, 971 (Paper I)
\bibitem[Busso et al. (1999)]{bu99} Busso, M., Gallino, R., \& Wasserburg, G.J. 1999, ARA\&A, 37, 239
\bibitem[Busso et al. (2007)]{bus07} Busso, M., Wasserburg, G.J., Nollett, K.M., \& Calandra, A. 2007, ApJ, 
671, 802
\bibitem[Cristallo (2006)]{cris06} Cristallo, S. 2006, PASP, 118, 1360
\bibitem[Cristallo et al. (2009)]{cris09} Cristallo, S., et al. 2009, ApJ, 696, 797 
\bibitem[Cristallo et al. (2010)]{cris10} Cristallo, S. et al. 2010, in preparation
\bibitem[Cunha et al. (2003)]{cun03} Cunha, K., Smith, V.V., Lambert, D.L., \& Hinkle, K.H. 2003, AJ, 126, 1305
\bibitem[Cunha (2005)]{cun05} Cunha, K., \& Smith, V.V. 2005, ApJ, 626, 425
\bibitem[Cunha et al. (2008)]{cun08} Cunha, K., Smith, V.V., \& Gibson, B. 2008, ApJ, 679, L17
\bibitem[Goriely (2000)]{gor00} Goriely, S., \& Mowlavi, N. 2000, A\&A, 362, 599
\bibitem[Grevesse (2007)]{gre07} Grevesse, N., Asplund, M., \& Sauval, A.J. 2007, Space Science Reviews, 130, 105
\bibitem[Guandalini (2008)]{gua08} Guandalini, R. 2008, AIP Conf. Proc., 1001, 339
\bibitem[Gustafsson et al. (2008)]{gus08} Gustafsson, B. et al. 2008, A\&A, 486, 951
\bibitem[Harris et al. (1987)]{har87} Harris, M.J., et al. 1987, ApJ, 316, 294
\bibitem[Johnson (1982)]{jo82} Johnson, H.R. 1982, ApJ, 260, 254 
\bibitem[Jorissen et al. (1992)]{jor92} Jorissen, A., Smith, V.V., \& Lambert, D.L. 1992, A\&A, 261, 164 (JSL)
\bibitem[Lambert et al. (1986)]{lam86} Lambert, D.L., Gustafsson, B., Eriksson, K, Hinkle, K.H. 1986, 
ApJS, 62, 373
\bibitem[Lugaro et al. (2004)]{lug04} Lugaro, M. et al. 2004, ApJ, 615, 934
\bibitem[Meynet (2000)]{mey00} Meynet, G., \& Arnould, M. 2000, A\&A, 335, 176
\bibitem[Nollet (2003]{no03} Nollett,  K.M., Busso, M., \& Wasserburg, G.J. 2003, ApJ, 582, 1036 
\bibitem[Otsuka et al. (2008)]{ots08} Otsuka, M., Izumiura, H., Tajitsu, A., \& Hyung, S. 2008, ApJ, 682, L108 
\bibitem[Palacios et al. (2005)]{pal05} Palacios, A., Arnould, M., \& Meynet, G. 2005, A\&A, 443, 243
\bibitem[Renda et al. (2004)]{ren04} Renda, A., et al. 2004, MNRAS, 354, 575
\bibitem[Smith (1990)]{smi90} Smith, V.V., \& Lambert, D.L. 1990, ApJS, 72, 387
\bibitem[Stanclifo (2005)]{stan05} Stancliffe, R.J., et al. 2005, MNRAS, 360, 375
\bibitem[Uttenthaler et al. (2008)]{ute08} Uttenthaler, S. et al. 2008, ApJ, 682, 509
\bibitem[Werner et al. (2005)]{wer05} Werner, K., Rauch, T., \& Kruk, J.W. 2005, A\&A, 433, 641
\bibitem[Woosley (1988)]{w88} Woosley, S.E., \& Haxton, W.C. 1988, Nature, 334, 5
\bibitem[Zamora (2009)]{za09} Zamora, O. 2009, Ph. D. Thesis, University of Granada


\end{thebibliography}
\end{document}